\documentclass[acmsmall]{acmart}
\pdfoutput=1
\usepackage{booktabs}
\usepackage{subfigure}
\usepackage{caption}
\usepackage{enumitem}
\usepackage{balance}
\usepackage{multirow}
\usepackage[table,xcdraw]{colortbl}

\newcommand{\effg}{\ensuremath{\mathrm{effG}}}
\newcommand{\effu}{\ensuremath{\mathrm{effCO}}}
\newcommand{\effo}{\ensuremath{\mathrm{effO}}}

\newcommand{\cvac}{CV-ABAC$\mathrm{_G}$}

\newcommand{\effgcv}{\ensuremath{\mathrm{effG}}}
\newcommand{\effucv}{\ensuremath{\mathrm{effCO}}}
\newcommand{\effocv}{\ensuremath{\mathrm{effO}}}
\newcommand{\Grcv}{\ensuremath{\mathrm{G}}}
\newcommand{\rangecv}{\ensuremath{\mathrm{Range(att)}}}
\newcommand{\dirugcv}{\ensuremath{\mathrm{directG}}\xspace}

\AtBeginDocument{%
  \providecommand\BibTeX{{%
    \normalfont B\kern-0.5em{\scshape i\kern-0.25em b}\kern-0.8em\TeX}}}

\setcopyright{none}
\renewcommand\footnotetextcopyrightpermission[1]{}
\begin{document}

 \title{Secure Cloud Assisted Smart Cars Using ABAC and Dynamic Groups}
 \title{Attribute Based Access Control with Dynamic Groups for Secure Cloud Assisted Smart Cars}
 \title[Secure Cloud Assisted Smart Cars Using Dynamic Groups and ABAC]{Secure Cloud Assisted Smart Cars Using Dynamic Groups and Attribute Based Access Control}

\author{Maanak Gupta}
\email{gmaanakg@yahoo.com}
\affiliation{%
  \institution{\\Department of Computer Science, Tennessee Tech. University}
  \city{Cookeville, Tennessee}
  \country{USA}}
\author{James Benson}
\email{james.benson@utsa.edu}
\author{Farhan Patwa}
\email{farhan.patwa@utsa.edu}
\author{Ravi Sandhu}
\email{ravi.sandhu@utsa.edu}
\affiliation{%
  \institution{\\Institute for Cyber Security, University of Texas at San Antonio}
  \city{San Antonio, Texas}
  \country{USA}}

\authorsaddresses{Most of this work was performed while the first author was a Postdoctoral Fellow at the University of Texas at San Antonio. This paper is an extended version of the research work first appeared in Gupta et al. \cite{Gupta:2019:DGA:3292006.3300048}. \\Authors' address : Maanak Gupta, Department of Computer Science, Tennessee Tech. University, Cookeville, Tennessee, USA, email: gmaanakg@yahoo.com; James Benson, Farhan Patwa and Ravi Sandhu, Department of Computer Science and Institute for Cyber Security, University of Texas at San Antonio, San Antonio, Texas, USA; emails: james.benson@utsa.edu, farhan.patwa@utsa.edu, ravi.sandhu@utsa.edu}

\renewcommand{\shortauthors}{Gupta et al.}

\begin{abstract}
Future smart cities and intelligent world will have connected vehicles and smart cars as its indispensable and most essential components. The communication and interaction among such connected entities in this vehicular internet of things (IoT) domain, which also involves smart traffic infrastructure, road-side sensors, restaurant with beacons, autonomous emergency vehicles, etc., offer innumerable real-time user applications and provide safer and pleasant driving experience to consumers. Having more than 100 million lines of code and hundreds of sensors, these connected vehicles (CVs) expose a large attack surface, which can be remotely compromised and exploited by malicious attackers. Security and privacy are serious concerns that impede the adoption of smart connected cars, which if not properly addressed will have grave implications with risk to human life and limb. In this research, we present a formalized dynamic groups and attribute-based access control (ABAC) model (referred as \cvac) for smart cars ecosystem, where the proposed model not only considers system wide attributes-based security policies but also takes into account the individual user privacy preferences for allowing or denying service notifications, alerts and operations to on-board resources.
Further, we introduce a novel notion of groups in vehicular IoT, which are dynamically assigned to moving entities like connected cars, based on their current GPS coordinates, speed or other attributes, to ensure relevance of location and time sensitive notification services to the consumers, to provide administrative benefits to manage large numbers of smart entities, and to enable attributes and alerts inheritance for fine-grained security authorization policies. We present proof of concept implementation of our model in AWS cloud platform demonstrating real-world uses cases along with performance metrics.
\end{abstract}

\maketitle

\section{Introduction}
Internet of Things (IoT) has become a dominant technology which has proliferated to different application domains including health-care, homes, industry, power-grid, to make lives smarter. It is predicted \cite{forbes} that the global IoT market will grow to \$457 Billion by year 2020, attaining a compound annual growth  rate (CAGR) of 28.5\%. Automation is leading the world today, and with `things' around sensing and acting on their own or with a remote user command, has given humans to have anything accessible with a finger touch. Data generated by these smart devices unleash countless business opportunities and offer customer targeted services. IoT smart devices along with `infinite' capabilities of cloud computing are ideally matched with desirable synergy in current technology-oriented world, which has been often termed by researchers as cloud-enabled, cloud-centric or cloud-assisted IoT in literature \cite{bhatt2017access,bhatt2017wearable,6778179,7037764,gupta2018authorization}.

IoT is embraced by every industry with automobile manufacturers and transportation among the most aggressive. The global connected car market is expected to reach  USD 219.21 billion by 2025 \cite{conn-car-report} with a CAGR of 14.8\%. Vehicular IoT inherits intrinsic IoT characteristics but dynamic pairing, mobility of vehicles, real-time, location sensitivity are some features which separates it from common IoT applications. The vision of smart city incorporates intelligent transportation where connected vehicles (CVs) can `talk' to each other (V2V) and exchange information to ensure driver safety and offer location-based services. These intelligent vehicles can also interact with smart roadside infrastructure (V2I), with pedestrian on road (V2H) or send data to the central cloud for processing and use. Basic safety messages (BSMs) are exchanged among moving entities using commonly used WiFi like secure and reliable Dedicated Short Range Communication (DSRC) protocol. Vehicles can receive speed limit notification, flash flood alerts or deer threat warnings on car dashboard or with a seat vibration. A car will receive information about nearby parking garages, restaurant offers or remote engine monitoring by authorized mechanic with nearby repair facility and discounts updating automatically. These services will provide pleasant travel experience to drivers and unleash business potential in this intelligent transportation domain. Smart internet connected vehicles embed softwares having more than 100 million lines of code to control critical systems and functionality, with plethora of sensors and electronic control units (ECUs) on board generating huge amounts of data so these vehicles are often termed as `datacenter on wheels'.

The conventionally isolated and disconnected vehicles now get exposed to external environment and internet, they become vulnerable to cyber attacks. Common security vulnerabilities including buffer overflow, malware, privilege escalation, and trojans etc. can be easily exploited in connected vehicles. Other potential threats include untrustworthy or fake messages from smart objects, malicious software injection, data privacy, ECU hacking and control, and spoofing connected vehicle sensor. With broad attack surface exposed via air-bag ECU, On-Board Diagnostics (OBD) port, USB, Bluetooth, remote key, and tire-pressure monitoring system etc. these attacks have become much easier to orchestrate. In-vehicle Controller Area Network (CAN) bus also needs security to protect message exchange among ECUs. Further, communication with external networks including cellular, WiFi and insecure public networks of gas stations, toll roads, service garages, or after-market dongles are a big threat to connected vehicles security. Cyber incidents including Jeep \cite{jeep} and Tesla Model X \cite{tesla} hacks where engine was stopped and steering remotely controlled demonstrate security vulnerabilities. Uber self driving car accident \cite{uber} in 2018 was due to a disabled emergency stop system. It could have been a result of remote adversary disabling the system, thereby compromising their complete fleet on the ground. Smart car incidents have serious implications as they can even result in loss of human life.

Access control \cite{sandhu1994access,sandhu1996role,ferraiolo2001proposed} mechanisms are widely used to restrict unauthorized access to resources and secure communication among entities. Attribute-based access control (ABAC) \cite{hu2015attribute,jin2012unified} provides finer granularity and offers flexibility in distributed multi-entity communication scenarios, which considers characteristics of participating entities along with system and environment properties to determine access decision. Smart cars ecosystem involves dynamic interaction and message exchange among connected objects, which must be authorized. It is necessary that only legitimate entities are allowed to control on-board sensors, data messages and send notifications. Further, user-centric privacy requires that end-users and customers can control what kind of alerts they want to receive, what advertisements they are interested or who can access their car's critical sensors, etc. This paper focuses on the access control needs in connected smart cars and proposes an attribute-based access control model for connected vehicles\footnote{In this research paper, the terms smart cars and connected vehicles have been used interchangeably which also subsumes autonomous vehicles.} ecosystem, referred as \cvac. Our solution considers the attributes of moving entities like current location, speed etc. to dynamically assign them to various groups (predefined by smart city administration), for implementing attributes-based security policies, and also incorporate user-specific privacy preferences for ensuring relevance of notifications service in constantly changing and mobile smart cars ecosystem. 
We implemented a prototype of our model as an external authorization engine hooked into the widely used AWS (Amazon Web Services) cloud platform \cite{aws}.

Rest of the paper is organized as follows. Section \ref{related} discusses related work and reviews the extended access control architecture (E-ACO) recently proposed for vehicular IoT environment. Section \ref{sec auth} highlights multi-layered authorization requirements and emphasize the need to introduce dynamic groups in smart cars applications. Section \ref{model} presents and formalizes our proposed groups and attribute-based access control model (\cvac) for connected vehicles ecosystem. Section \ref{use-case} provides AWS cloud implementation of dynamic groups assignment of moving entities based on attributes and discusses our external policy decision and enforcement engine for security policies along with detailed performance metrics and evaluation. Section \ref{summary} summarizes our work.


\section{Related Work} \label{related}

Vehicular IoT and smart cars involve dynamic communications and data exchange which requires access controls to restrict within authorized entities. In this section, we first discuss a recently proposed extended access control architecture (E-ACO) which focuses on access control requirements in connected vehicles. We also highlight some important work done by government and private agencies to gauge cyber risks and security measures in smart vehicles.

\subsection{Extended Access Control Oriented Architecture}

\begin{figure*}[t]
  \centering
  \subfigure[Four Layered Architecture]{\includegraphics[scale=0.34]{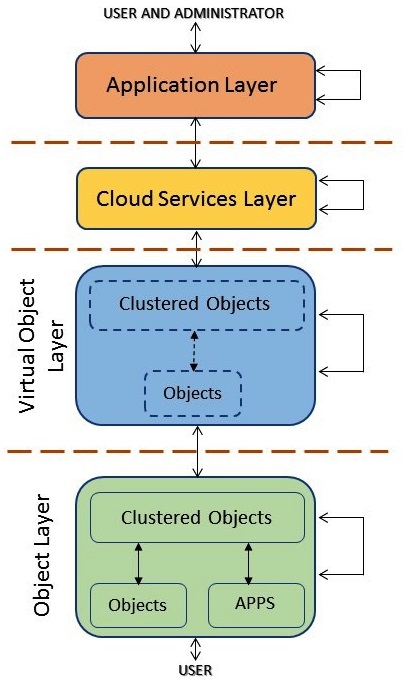}\label{fig eaco}}\quad
  \subfigure[Vehicular IoT Components in E-ACO Layers]{\includegraphics[scale=0.34]{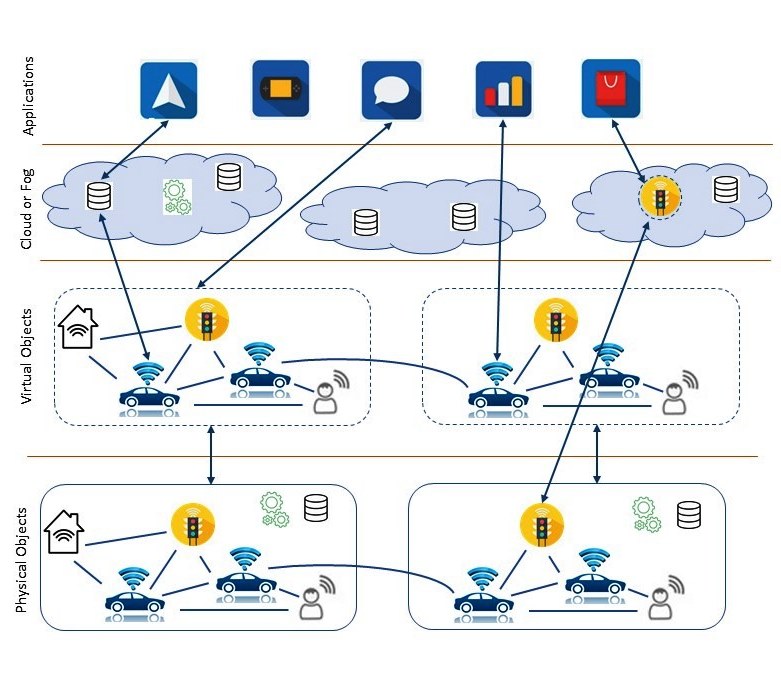}\label{fig eaco-b}}
  \caption{Extended Access Control Oriented Architecture \cite{gupta2018authorization}}
  \label{fig eacoc}
\end{figure*}

Several IoT architectures with multi-layer stack have been discussed in literature \cite{al2015internet,atzori2010internet,6984170,gubbi2013internet,alshehri2016access}. 
 Alsehri and Sandhu \cite{alshehri2016access} recently presented a general IoT architecture which includes virtual objects \cite{nitti2016virtual} and cloud as two middleware layers. Virtual objects resolve IoT issues of heterogeneity and connectivity whereas on-demand capabilities of cloud are in cloud service layer. Gupta and Sandhu \cite{gupta2018authorization,gupta2018s&p,gupta2018dissertation} extended this IoT architecture for specific vehicular IoT and connected vehicles domain. This extended access control architecture (E-ACO), shown in Figure \ref{fig eacoc}, introduced clustered objects (like smart cars and traffic lights) which are objects with multiple individual sensors. Also, these clustered objects have applications (for example, lane departure or safety warning system in cars) installed on board, which is usually not the case in general IoT realm. Shown in Figure \ref{fig eaco}, E-ACO is a four-layered architecture defined as follows:

\noindent
\textbf{Object Layer :} This is the bottom most layer which represents real physical clustered objects and sensors along with applications installed on them. In-vehicle communication at this layer is mainly supported by Ethernet and CAN technologies, whereas communication across clustered objects is done using DSRC (used for BSM exchange in V2V or V2X communication), WiFi, or LTE etc. 
It should be noted that each layer in E-ACO architecture interacts within itself and with entities in adjacent layers, as marked by arrows in the figure. Therefore, object layer will interact with users at the bottom and virtual object layer above it.

\noindent
\textbf{Virtual Object Layer :} This layer acts as an intermediate between cloud services and physical layer, which offers necessary abstraction by creating cyber entities for physical objects in object layer. In connected vehicles domain, where cars move across different terrains where internet connectivity can be an issue, it is important to have cyber entities which maintain the state of the corresponding physical object as best known and to be updated when connectivity is restored. When two sensors s$_\mathrm{1}$ and s$_\mathrm{2}$ across different vehicles interact with each other, the order of communication using virtual objects will follow s$_\mathrm{1}$ to vs$_\mathrm{1}$ (virtual entity of sensor s$_\mathrm{1}$), vs$_\mathrm{1}$ to vs$_\mathrm{2}$ and vs$_\mathrm{2}$ to physical sensor s$_\mathrm{2}$.

\noindent
\textbf{Cloud Services and Application Layer :} As applications use cloud services, therefore these two layers are discussed together. On-board sensors generate data which is stored and processed by cloud services, which is used by applications to offer services to end-users. Cyber entities of physical objects can be created in cloud layer which provides a persistent state information of objects. It is important to mention that central cloud may incur latency and bandwidth issues in time-sensitive applications which can be resolved by introducing edge or fog computing infrastructure.

Figure \ref{fig eaco-b} shows an instance of vehicular IoT with physical objects (car,
traffic light or beacons) along with their cyber counterparts in virtual objects layer,
and other E-ACO layers. It can be noted that physical objects communicate with their virtual objects, and applications are accessing data
through cloud which is pushed by virtual entity of an object. Storage
and processing icons at object layer symbolizes road-side infrastructures which can help to store data from smart vehicles and filter it before
pushing data to cloud to save bandwidth. Virtual objects can be created at both fog and
central cloud to satisfy different applications and use-cases.

\subsection{Relevant Background and Technologies}

Smart cars and associated applications are still in early stages but involve some established technologies. Vehicular Ad-hoc Networks (VANETs) \cite{vanets} have been discussed which support vehicle to vehicle and infrastructure communication for user services. In VANETS, moving cars and infrastructure act as network nodes to provide storage, computation and other services. This concept is further extended with the inclusion of cloud computing. Vehicular Clouds (VC) \cite{10.1007/978-3-642-17994-5_1,gerla2014internet,olariu2011taking} were proposed to integrate VANETs and cloud, to offer on-the fly edge/cloud platform to cars and applications by utilizing on-board resources. VCs are relevant in smart cars real-time and location-centric applications and services, which are otherwise impractical due to latency and bandwidth issues of central cloud. Several VC architectures have been discussed including stationary, fixed infrastructure or dynamic \cite{hussain2012rethinking,whaiduzzaman2014survey}.

Cyber threats to connected vehicles are very serious concerns. Government agencies and private sectors are well aware of the risks involved and want to ensure that no open doors are left to orchestrate attacks before wide adoption. The US Department of Transportation (USDOT) has invested in Intelligent Transportation System (ITS) \cite{its} which has connected vehicles as an important component with aim to reduce accidental fatalities. Cyber security is a key area and along with National Highway Traffic Safety Administration (NHTSA), it has released cyber-security guidelines \cite{nhtsa-1,nhtsa-2}. Security Credential Management System (SCMS) \cite{its-scms} is proposed as DSRC message security solution in vehicle-to-vehicle (V2V) and vehicle-to-infrastructure (V2I) communication. It uses Public Key Infrastructure (PKI)-based approach to enable trusted interaction where a certificate authority issued certificate is attached to each BSM \cite{its-1} to ensure vehicle trustworthiness. US Government Accountability Office (GAO) \cite{gao} have widely discussed vulnerabilities and attack surfaces in smart vehicles, and also proposed solutions to prevent such threats. European Union Agency for Network and Information Security (ENISA) also studied critical assets and threats in smart cars together with security mechanisms to mitigate them \cite{enisa}. Cooperative Intelligent Transport Systems (C-ITS) for European Union \cite{c-its1,c-its2} has defined a PKI-based trust model to ensure authenticity and integrity of vehicle messages.

Homomorphic encryption based security solutions and protocols have been extensively discussed to provide location proximity \cite{hallgren2015innercircle,narayanan2011location,zhong2007louis} which can help to provide location based services without sharing the exact coordinates of drivers. These approaches can be used and complement our proposed \cvac~ model to resolve the privacy concerns of end users.

Access controls are widely used in computer systems to restrict unauthorized access to resources.
Park et al \cite{park2011acon,park2011user} proposed an activity centric access control model for social networks which considers user privacy policies in access decision. \cvac~model is inspired from this work besides being a pure ABAC model with dynamic groups which are pertinent in smart cars ecosystem.

\section{Access Control Needs in Connected Smart Cars} \label{sec auth}

Smart cars expose the conventionally isolated car systems to external environment via internet. The dynamic and short-lived real time V2V and V2I interaction with entities in and around connected vehicle needs to ensure message confidentiality and integrity, as also protection of on-board resources from adversaries. This section provides an overview of access control requirements and underlines the need for dynamic groups in smart vehicles IoT domain. 

\subsection{Multi-Layer Security Requirements and User Privacy Preferences}

Broad attack surface exposed by connected vehicles is the first entry point to in-vehicle critical systems. We believe that two level access control policies are the minimum essential to protect the external interfaces and internal ECU communication. Access control for external environment will protect on-board sensors, applications and user personal data from unauthorized access by entities including vehicles, applications, masquerading remote mechanics or other adversaries. Over-the air firmware update needs to be checked and must be allowed only from authorized sources. An attacker even if successful in passing through the first check point, must be restricted at the in-vehicle level, which secures overwrite and control of critical units (engine, brakes, telematics etc.) from adversaries. Vehicles exchange BSMs which raises an important question about trust. It must be ensured that information received is correct and from a trusted party, before being used by on-vehicle applications. Applications access sensors within and outside the car, which must be authorized, for example, a lane departure warning system accessing tire sensors must be checked to prevent a spoofed application reading vehicle movements. A passenger accessing infotainment (information and entertainment) systems of the car via Bluetooth or using his smartphone inside car must also be authorized. Proper and resilient isolation is needed to protect critical vehicle systems from being compromised through exposed entry points.

Smart cars location-based services enable notifications and alerts to vehicles. A user must be allowed to set his personal preferences whether he wants to receive advertisements from certain sources or filter out which ones are acceptable. For instance, a user may not want to receive restaurant notifications but is interested in flash-flood warnings. Further, more fine grained policies may be defined by a user, for example, a driver only wants notifications from cheesecake factory and between 8 to 10 pm. System wide policy, like a speed warning to all over-speeding vehicles or a policy of who can control speed of autonomous car are needed.

Data protection in cloud is critical due to frequent occurrence of data breaches. Big Data access control  \cite{Gupta:2017:MAF:3078861.3084173,gupta2018attribute,Gupta2017,gupta2017poster} is essential when user privacy has to be ensured and unauthorized disclosure is not allowed. Cross cloud trust models are needed to allow data access when mechanic application in private cloud reads data in car-manufacturer cloud. Physical tampering of vehicle ECUs, OBD and sensors also require parameter protection but is out of scope for this paper.

\subsection{Relevance of Dynamic Groups in Mobile Vehicular IoT}
\begin{figure*}[t]
  \centering
  \subfigure[Smart City with Geographical Location Groups]{\includegraphics[scale=0.35]{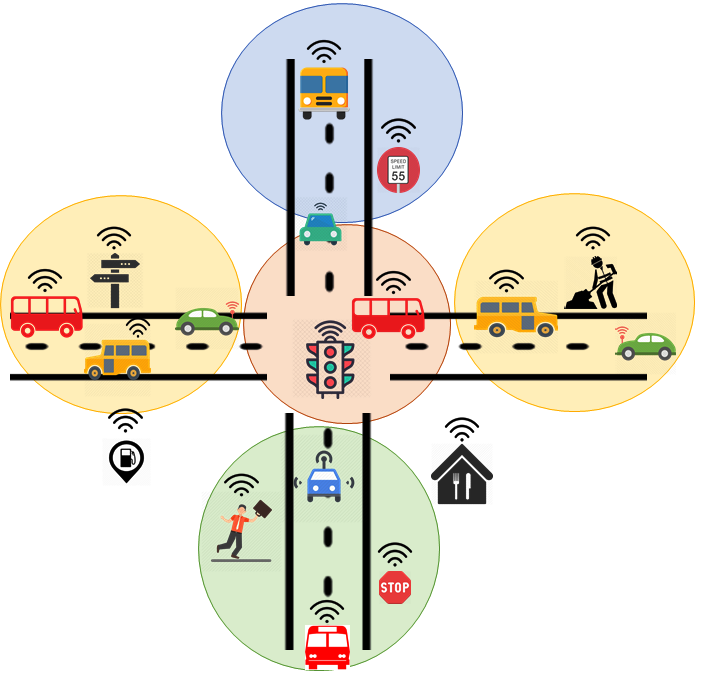} \label{fig groups}}\quad
  \subfigure[Example Groups Hierarchy]{\includegraphics[scale=0.35]{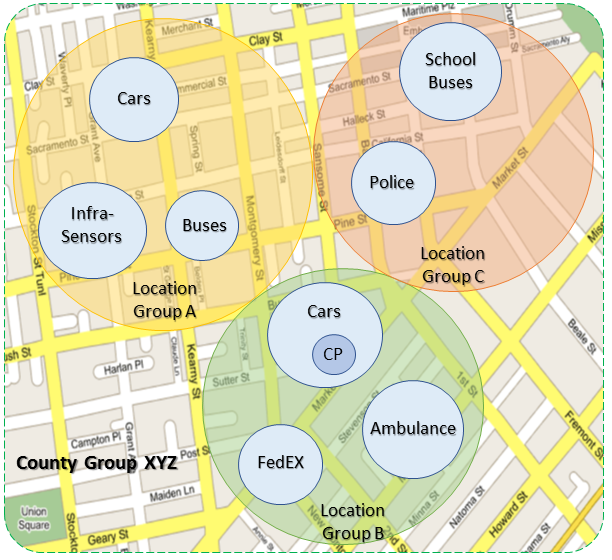} \label{fig gh}}
  \caption{Representative Groups in Connected Smart Cars Environment}
\end{figure*}

Most smart cars applications and service requests from drivers are location specific and time sensitive. For example, a driver might want to get warning signals when traveling near a blind spot, in school zone or pedestrians crossing road. Further, notifications sent to drivers are short-lived and mostly pertinent around current GPS coordinates. A gas discount notification from a nearby station, an accident warning two blocks away or ice on the bridge, are some example where alerts are sent to all vehicles in the area. Therefore, we believe that dynamically categorizing connected vehicles into location groups will be helpful for scoping the vehicles to be notified instead of a general broadcast and reduce administrative overheads, since single notification for the group will trigger alerts for all its members. Also, entities present at a location have certain characteristics (like stop sign warning, speed limit, deer-threat etc.) in common, which can be inherited by being a group member. Figure \ref{fig groups} represents how various smart entities can be separated into different location groups defined by appropriate authorities in a smart city system. These groups are dynamically assigned to connected vehicles based on their attributes, personal preferences, interests or current GPS coordinates as further elaborated in the model and implementation section discussed later.

Groups hierarchy can also exist, as shown in Figure \ref{fig gh}, with sub-groups within a larger parent group so as to reduce the number of vehicles to be notified. For instance, under location group, sub-groups can be created for cars, buses, police vehicles or ambulances, to enable targeted alerts to ambulances or police vehicle sub-groups defined within the location group. Groups can be defined based on services, for example, a group of cars within the car parent group which take part in car-pooling (CP) service or those which want to receive gas station offers. Group hierarchy \cite{servos2014hgabac,gupta2016mathrm} also enables attributes inheritance from parent to child groups. It helps in easy propagation and administration of alerts (like flash flood, deer threat or ice on road), where an alert generated at higher level of hierarchy (like a location group) is automatically trickled to all its children groups.



\section{Access Control Model for Connected Vehicles Ecosystem} \label{model}

Dynamic communication and data exchange among entities in connected vehicles ecosystem require multi-layer access control policies, which are managed centrally and also driven by individual user preferences. Therefore, an access control model must incorporate all such user and system requirements and offer fine-grained authorization solutions. In this section, we will discuss and formally define our proposed connected vehicle attribute-based access control model with dynamic groups, which we refer as \cvac.
\subsection{\cvac~Model Components}

\begin{figure}[t]
\centering
\includegraphics[scale=.50]{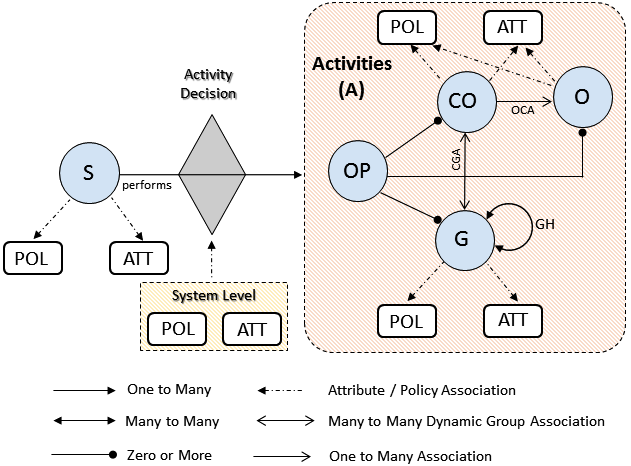}
\caption{A Conceptual \cvac~Model}
\label{fig model}
\end{figure}

The conceptual \cvac~model is shown in Figure \ref{fig model} with formal definitions summarized in Table \ref{tab 1} and continued in Table \ref{tab-ch4-2}. The basic model has following components: Sources (S), Clustered Objects (CO), Objects in clustered objects (O), Groups (G), Operations (OP), Activities (A), Authorization Policies (POL), and Attributes (ATT).

\noindent
\textbf{Sources (S):} These entities initiate activities (explained below) on various smart objects, groups and applications in the ecosystem. A source can be a user, an application, administrator, sensor, hand-held device, clustered object (such as a connected car), or a group defined in the system. For instance, in case of flash flood or deer threat warning, activity source is police or city department triggering an alert to all vehicles in the area. Similarly, car mechanic is a source, when he tries to access data from on-board engine sensor in the car using his remote cloud based application. A restaurant or gas-station issuing coupons are also considered as source.


\noindent
\textbf{Clustered Objects (CO):} Clustered objects are particularly relevant in case of connected vehicles, traffic lights or smart devices held by humans as they have multiple sensors and actuators. A smart car with on-board sensors, ECUs (like tire pressure, lane departure, or engine control) and applications is a clustered object. These smart entities interact and exchange data among themselves and with others such as requestor source, applications or cloud. An important reason to incorporate clustered objects is to reflect cross-vehicle and intra-vehicle communication. The fact that two smart vehicles can exchange basic safety messages (BSM) with each other shows clustered object communication.

\noindent
\textbf{Objects in clustered objects (O):} These are individual sensors, ECUs and applications installed in clustered objects. Objects in smart cars include sensors for internal state of the vehicle, e.g., engine diagnostics, emission control, cabin monitoring system, as well as sensors for external environment such as cameras, temperature, rain, etc. Control commands can directly be issued to these objects, and data can be read remotely. Applications (like lane departure warning system) on board can also access data from these objects to provide alerts to driver or to a remote service provider.

\noindent
\textbf{Groups (G):} A group is a logical collection of clustered objects with similar characteristics or requirements. With these groups, subset of COs can be sent relevant notification and also attributes can be assigned to group members. Some groups which can be defined smart vehicles ecosystem include location specific groups, service specific groups (like car-pooling, gas station promotions etc.) or vehicle type (a group of cars, buses etc.). Group hierarchy (GH) also exists which enables attributes and policies inheritance from parent to children groups. For simplicity, we require that a vehicle or CO can be direct member of only one group at same hierarchy level. For example, a car can be in either location A or B group and but not both. Such restriction helps in managing attributes inheritance and enhances the usability of our model.


\noindent
\textbf{Operations (OP):} These are actions which can be performed against clustered objects, individual objects or groups. Examples include: a mechanic performing read, write or control operations on engine ECU, a restaurant triggering notifications to vehicles in location A group. Operations also include administrative actions like creating or updating attributes or policies for COs, objects and groups, which are usually performed by system/security administrators.

\noindent
\textbf{Activities (A):} Activities encompass both operational and administrative activities which are performed by various sources in the system. An activity can have one or many atomic operations (OP) involved and will need authorization policies, which can be user privacy preferences, system defined or both, to allow or deny an activity. For example, a car pooling notification activity generated by a requestor (source) will be broadcast to all relevant vehicles in the locations nearby using location groups, however individual drivers must also receive or respond to that request based on individual preferences. A driver may not want to car-pool the requestor because of poor rating or because he is not going to the destination the requestor asked for. Therefore, an activity can involve multiple set of policies defined at different levels which must be evaluated, in car-pooling case a policy is set to determine cars to be notified and then driver personal preferences. We have primarily divided these smart car activities into following categories.
\noindent
\begin{itemize}[leftmargin=10pt]
  \item Service Requests: These are activities initiated by entities or users (via applications). For instance, a vehicle break-down initiates a service request to other vehicles around, or a user using a smartphone initiates a car-pooling requests for a destination to cars which are available or have opted in for the service.
  \item Administration: These activities perform administrative operations in system which include changing policies and attributes of entities or determining the group hierarchy. It also defines the scope of groups, how user privacy preferences are used, or how vehicles are determined to be a member of a group etc. 
  \item Notifications: These are group centric activities where all members are notified for any updates about the group (like speed limit or deer threat notifications in location A) or for locations-based marketing promotions by parking lots or restaurants.
  \item Control and Usage: These activities include simple read, write or control operations performed remotely or within a vehicle. Over the air updates issued by manufacturer or turning on car climate control using a smart key are remote activities whereas a passenger accessing infotainment system using smartphone and on-board car applications reading car camera are local.
\end{itemize}


\noindent
\textbf{Authorization Policies and Attributes:} \cvac~model incorporates individual user privacy controls for different entities by managing authorization policies and entity attributes. As shown in Figure \ref{fig model}, policy of sources include personal preferences, whereas attributes reflect characteristics like name, age or gender. Policies can be defined for clustered objects, for instance, a USB can be plugged only by car owner, or which mechanic can access an on-board sensor. Attributes of a car include GPS coordinates, speed, heading direction, and vehicle size. Groups also set policies and attributes for themselves, for example, car pooling group policy of who can be its member. Similarly, system wide policies are also considered, for instance, policy to determine which groups will be sent information when a request comes from a source, or policy to change group hierarchy. Policies also include attributes of entities involved in an activity. A CO can inherit attributes from dynamically assigned groups which will change as the CO leaves old group and adds to new group.

It should be noted that attributes of entities change more often than system wide or individual policies. Attributes are more dynamic in nature which are added or removed with the movement of vehicles or change in surroundings, like GPS coordinates or temperature. Policies once set by administrators or users are more static and only the attributes which comprise the policy change the outcome of a policy but the policy definition remains relatively fixed. For instance, a user policy could state that `Send restaurant notifications only from Cheesecake factory'. In such case, only attribute name of the restaurant sending the notification will be checked and if it is equal to Cheesecake factory will be able to advertise to that user. Dynamic policies are also possible, for instance, a policy may state that police vans in locations groups A and B are notified in case of emergency, but, in case of a bigger threat this policy can be changed or overwritten with police vans in groups A, B C and D. Our proposed model assumes that no policies or attributes are changed during an activity evaluation process.

Some activities in the system will need multi-level policy evaluation and may also include user privacy preferences before making a decision. For instance, a user must be allowed to decide if he wants to share data from car sensors or whether wants to get marketing advertisements. Each activity will evaluate required system and user policies to make final decision.
\begin{table}
\centering
\caption{Formal \cvac~Model Definitions for Connected Vehicles Ecosystem}
\label{tab 1}
{%
\begin{tabular}{@{}llllll@{}}
\toprule
\multicolumn{6}{l}{\textbf{Basic Sets and  Functions}}\\
\multicolumn{6}{l}{\begin{tabular}{@{}l@{}}-- S, CO, O, G, OP are finite sets of sources, clustered objects, objects, groups and operations \\\;\; respectively [blue circles in Figure \ref{fig model}].\end{tabular}}                              \\
\multicolumn{6}{l}{-- A is a finite set of activities which can be performed in system.}                                                      \\
\multicolumn{6}{l}{-- ATT is a finite set of attributes associated with S, CO, O, G and system-wide.}                                                      \\
\multicolumn{6}{l}{-- For each attribute att in ATT,  Range(att) is a finite set of atomic values.}
\\
\multicolumn{6}{l}{-- attType: ATT = \{set, atomic\}, defines attributes to be set or atomic valued.}                                                      \\
\multicolumn{6}{l}{\begin{tabular}{@{}l@{}}-- Each attribute att in ATT maps entities in S, CO, O, G to attribute values. Formally,\\\;\; att : S $\cup$ CO $\cup$ O $\cup$ G $\cup$ \{system-wide\} $\rightarrow$\; $ \begin{cases} \mathrm{{Range(att)}}~\cup \{\bot\} & \text{if attType(att) = atomic}\\ \mathrm{2^{Range(att)}} & \text{if attType(att) = set} \\ \end{cases}$ \end{tabular}}
\\
\multicolumn{6}{l}{-- POL is a finite set of authorization policies associated with individual S, CO, O, G.}                                                      \\
\multicolumn{6}{l}{\begin{tabular}{@{}l@{}}-- directG : $ \mathrm{CO \rightarrow {G}} $,  mapping each clustered object to a system group, \\\;\; equivalently CGA $ \subseteq \mathrm{CO} \times \mathrm{G}. $\end{tabular}}
\\
\multicolumn{6}{l}{\begin{tabular}{@{}l@{}}-- parentCO : $ \mathrm{O \rightarrow CO} $,  mapping each object to a clustered object,\\\;\; equivalently OCA $ \subseteq \mathrm{O} \times \mathrm{CO}. $\end{tabular}}
\\
\multicolumn{6}{l}{\begin{tabular}{@{}l@{}}-- $\mathrm{GH \subseteq G~\times}$ G,  a partial order relation $\succeq_{\textit{g}}$ on G. \\\;\; Equivalently, parentG : $\mathrm{G \rightarrow 2^{G}}$, mapping group to a set of parent groups in hierarchy.\end{tabular}}                                                                                                          \\
\\
\multicolumn{1}{l}{\textbf{Effective Attributes of Groups, Clustered Objects and Objects (Derived Functions)}}                                                                 \\
\multicolumn{6}{l}{\begin{tabular}[c]{@{}l@{}}-- For each attribute $\mathrm{att}$ in ATT such that attType(att) = set: \\ $ \;\;\;\; \bullet \;\;$ $\mathrm{\effgcv_{att}}$ : \Grcv~$\rightarrow {2^{\rangecv}}$, defined as $\mathrm{\effgcv_{att}(g_i)}$ = $\mathrm{att(g_i)}$ $\cup$  $ (\mathrm{\bigcup\limits_{ g \; \in\; \{g_j | g_i \; \succeq_{\textit{g}}\; g_j\} } \effgcv_{att}(g))}$. \end{tabular}} \\
\multicolumn{6}{l}{\begin{tabular}[c]{@{}l@{}}$ \;\;\;\; \bullet \;\;$ $\mathrm{\effucv_{att}}$ : CO~$\rightarrow {2^{\rangecv}}$, defined as $\mathrm{\effucv_{att}(co)}$ = $\mathrm{att(co)}$ $\cup$  $ \mathrm{ \effgcv_{att}(directG(co))}$. \end{tabular}} \\
\multicolumn{6}{l}{\begin{tabular}[c]{@{}l@{}}$ \;\;\;\; \bullet \;\;$ $\mathrm{\effocv_{att}}$ : O~$\rightarrow {2^{\rangecv}}$, defined as $\mathrm{\effocv_{att}(o)}$ = $\mathrm{att(o)}$ $\cup$  $\mathrm{ \effucv_{att}(parentCO(o))}$. \end{tabular}} \\
\\
\multicolumn{6}{l}{\begin{tabular}[c]{@{}l@{}}-- For each attribute $\mathrm{att}$ in ATT such that attType(att) = atomic: \\ $ \;\;\;\; \bullet \;\;$ $\mathrm{\effgcv_{att}}$ : \Grcv~$\rightarrow {{\rangecv}} \cup \{\bot\}$, \\\;\;\;\;\;\;\;\;\;defined as $\mathrm{\effgcv_{att}(g_i)}$ = $ \begin{cases} \mathrm{att(g_i)} & \text{if $\forall \mathrm{g'} \in \mathrm{parentG(g_i)}.~ \mathrm{\effgcv_{att}(g')} = \bot$}\\ \mathrm{\effgcv_{att}(g')} & \text{if $\exists~\mathrm{parentG(g_i)}.~ \mathrm{\effgcv_{att}(parentG(g_i))} \neq \bot$ then select}\\ & \text{parent $\mathrm{g'}$ with $\mathrm{\effgcv_{att}(g')} \neq \bot$ updated most recently.}\end{cases}$ \end{tabular}} \\
\multicolumn{6}{l}{\begin{tabular}[c]{@{}l@{}}$ \;\;\;\; \bullet \;\;$ $\mathrm{\effucv_{att}}$ : CO~$\rightarrow {{\rangecv}} \cup \{\bot\}$, \\\;\;\;\;\;\;\;\;\;defined as $\mathrm{\effucv_{att}(co)}$ = $ \begin{cases} \mathrm{att(co)} & \text{if $\mathrm{\effgcv_{att}(directG(co))} = \bot$}\\ \mathrm{\effgcv_{att}(directG(co))} & \text{otherwise}\\ \end{cases}$ \end{tabular}} \\
\multicolumn{6}{l}{\begin{tabular}[c]{@{}l@{}}$ \;\;\;\; \bullet \;\;$ $\mathrm{\effocv_{att}}$ : O~$\rightarrow {{\rangecv}} \cup \{\bot\}$,\\\;\;\;\;\;\;\;\;\; defined as $\mathrm{\effocv_{att}(o)}$ = $ \begin{cases} \mathrm{att(o)} & \text{if $\mathrm{\effucv_{att}(parentCO(o))} = \bot$}\\ \mathrm{\effucv_{att}(parentCO(o))} & \text{otherwise}\\ \end{cases}$ \end{tabular}} \\
\bottomrule
\end{tabular}}
\end{table}
\begin{table}[t!]
\centering
\caption{Formal \cvac~Model Definitions for Connected Vehicles Ecosystem (Continued)}
\label{tab-ch4-2}
\resizebox*{!}{\totalheight}{
\begin{tabular}{@{}llllll@{}}
\toprule
\multicolumn{2}{l}{ \textbf{Authorization Functions (Policies)}}                                                                                                          \\
\multicolumn{6}{l}{\begin{tabular}[c]{@{}l@{}}-- {Authorization Function:} For each op $\in$ OP, $\mathrm{Auth_{op}(s:S, ob:CO \cup O \cup G)}$ is a propositional \\\;\; logic formula  returning true or false,  which is defined  using the following policy language: \end{tabular}}\\
\multicolumn{6}{l}{\begin{tabular}[c]{@{}l@{}}$\;\;\;\; \bullet \;\; \mathrm{ \alpha \Coloneqq \alpha \land \alpha \;|\; \alpha \vee \alpha \;|\; (\alpha) \;|\; \neg \alpha \;|\; \exists\; x \in set.\alpha  \;|\; \forall\; x \in set.\alpha \;|\;  set \bigtriangleup set \;|\; } $ $ \mathrm{atomic \in set \;|\;}$ \\\;\;\;\;\;\;\; $ \mathrm{atomic \notin set}$ \end{tabular}} \\

\multicolumn{6}{l}{$ \;\;\;\; \bullet \;\; \bigtriangleup \Coloneqq  \; \subset \;|\; \subseteq \;|\; \nsubseteq \;|\; \cap \;|\; \cup $}\\
\multicolumn{6}{l}{\;\;\;\;$\bullet$\;\;for $\mathrm{att}$ $\in $ ATT, i $\in$ S $\cup$ CO $\cup$ O $\cup$ G $\cup$ \{system-wide\}, attType(att) = set :}\\
\multicolumn{6}{l}{$ \;\;\;\; \;\;\;\; \mathrm{  set \Coloneqq {eff}_{att}(i) \;|\; {att}(i) }$}\\

\multicolumn{6}{l}{\;\;\;\;$\bullet$\;\;for $\mathrm{att}$ $\in $ ATT, i $\in$ S $\cup$ CO $\cup$ O $\cup$ G $\cup$ \{system-wide\}, attType(att) = atomic :}\\
\multicolumn{6}{l}{$ \;\;\;\; \;\;\;\; \mathrm{ atomic \Coloneqq {eff}_{att}(i) \;|\; {att}(i) \;|\; value }$}\\
\midrule
\multicolumn{2}{l}{ \textbf{Authorization Decision}}                                                                                                          \\
\multicolumn{6}{l}{\begin{tabular}[c]{@{}l@{}}-- A source $\mathrm{s}$ $\in$ S is allowed to perform an activity a $\in$ A, stated as $\mathrm{Authorization(a:A, s:S)}$, \\\;\; if the required policies needed to allow  the activity are included and evaluated to make final \\\;\; decision. These multi-layer policies must be evaluated for individual operations ($\mathrm{op_i}$ $\in$ OP) \\\;\; to be performed by source s $\in$ S on relevant objects ($\mathrm{x_i}$ $\in$ CO $\cup$ O $\cup$ G). Formally, \\\;\;
$\mathrm{Authorization(a:A, s:S)}$ $\Rightarrow$ \\\;\; $\mathrm{Auth_{op_1}(s:S, x_1)}$, $\mathrm{Auth_{op_2}(s:S, x_2)}$, $\mathrm{Auth_{op_3}(s:S, x_3)}$, $\ldots$ $\ldots$ $\ldots$ $\ldots$ $\ldots$, $\mathrm{Auth_{op_n}(s:S, x_3)}$
\end{tabular}}
\\
\bottomrule
\end{tabular}}
\end{table}

\subsection{Formal Model Definitions} 
As shown in Table \ref{tab 1}, sources, clustered objects, objects and groups can be directly assigned values from the set of atomic values (denoted by Range(att)) for attribute att in set ATT. Each attribute can be a set or atomic valued, determined by attType function and based on its type, entities can be assigned a single value including null ($\bot$) for an atomic attribute, or multiple values for set-valued attribute from the attribute range. POL is the set of authorization policies defined in the system which will be defined below.

Clustered objects can be members of different groups, based on preferences and requirements. For example, a car is assigned to a location group based on its GPS coordinates. In our model, we assume that a clustered object can be directly assigned to only one group at same hierarchy level (specified by directG function). 
As we will discuss later that since groups inherit attributes from parent groups, assigning a clustered object to one parent group is sufficient to realize attributes inheritance. Smart cars have sensors and applications installed in them, which can also be accessed by different sources. Therefore, parentCO function determines the clustered object to which an object belongs, which is a one to many mapping i.e an object can only belong to one CO while a CO can have multiple objects. Further,
group hierarchy GH (shown as self loop on G), is defined using a partial order relation on G and denoted
by $\succeq_{\textit{g}}$, where $g_1$ $\succeq_{\textit{g}}$ $g_2$ signifies $g_1$ is child group of $g_2$ and $g_1$ inherits
all the attributes of $g_2$. Function parentG computes the set of parent groups in hierarchy for a child group.

The benefit to introduce groups is ease of administration where multiple attributes can be assigned or removed from member clustered objects with single administrative operation. Group hierarchy enables attributes inheritance from parent to child groups. Therefore, in case of set valued attributes, the effective attribute att of a group $\mathrm{g_i}$ (denoted by $\mathrm{\effg_{att}(g_i)}$) is the union of directly assigned values for attribute att and the effective values for att for all its parent groups in group hierarchy. This definition is well formed since $\succeq_{\textit{g}}$ is a partial order. For a maximal group $\mathrm{g_j}$ in this ordering, we have $\mathrm{\effg_{att}(g_j)}$ = att($\mathrm{g_j}$), giving base cases for this recursive definition. The effective attribute values of  clustered object for attribute att (stated as $\mathrm{\effu_{att}}$) will then be the directly assigned values for att and the effective attribute values of att for the group to which CO is directly assigned (by directG). Similarly, in addition to direct attributes, sensors in car can inherit attributes from the car itself (eg. make, model, location), $\mathrm{\effo_{att}}$ calculates these effective attributes of objects. For set valued attributes, union operation will be sufficient which is not true for atomic attributes. In case of groups, the most recently updated non-null attribute values in parent groups will overwrite the values of child group as defined in Table \ref{tab 1}. For example, if the most recent value updated in one of the parent groups for Deer\_Threat attribute is `ON', this value will trickle to the child group. It should be noted that overwriting with the most recently updated value in groups is one of the many approaches to inherit atomic attributes, but for the dynamic nature of smart cars ecosystem, we believe this is most appropriate. Clustered object inherits non-null atomic value from its direct parent group as stated by $\mathrm{\effu_{att}(co)}$ =  $\mathrm{ \effg_{att}(directG(co))}$. In case of objects, parent clustered object will overwrite non-null atomic attributes. For atomic attributes, if the parent(s) has null value for an attribute, the entity (group, clustered object or object) will retain its directly assigned value without any overwrite. As part of administrative work, attributes for entities must be carefully determined and allocated since inheritance may impact the attributes of other associated entities in the system.

Authorization functions are defined for each operation op $\in$ OP, which are policies defined in the system. POL is the set of all authorization functions, $\mathrm{Auth_{op}(s:S, ob:CO \cup O \cup G)}$, which specify the conditions under which
source s $\in$ S can execute operation op $\in$ OP on object ob $\in$ CO $\cup$ O $\cup$ G. Such policies include privacy preferences set by users for individual clustered object, objects and groups or can be system wide by security administrators. The conditions can be specified as propositional logic formula using policy language defined in Table \ref{tab-ch4-2}. Multiple policies must be satisfied before an activity is allowed to perform. Authorization function, $\mathrm{Authorization(a:A, s:S)}$, where an activity a $\in$ A is allowed by source s $\in$ S, specifies the system level, user privacy policies or other relevant policies returning true for an activity to succeed.

\begin{figure*}[t]
  \centering
  \subfigure[System Architecture]{\includegraphics[scale=0.45]{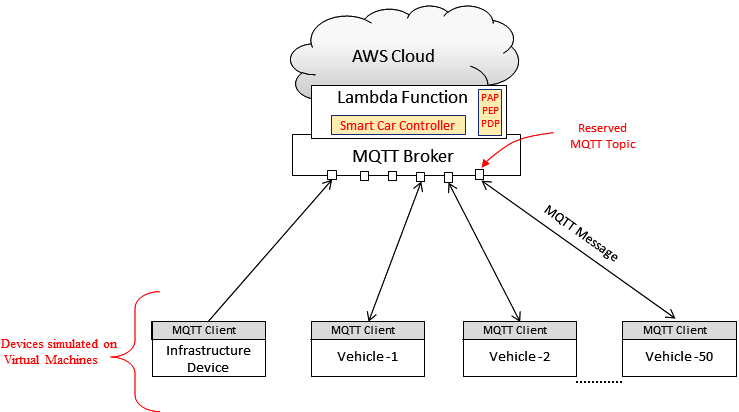} \label{fig-arch}}\quad
  \subfigure[AWS Components]{\includegraphics[scale=0.45]{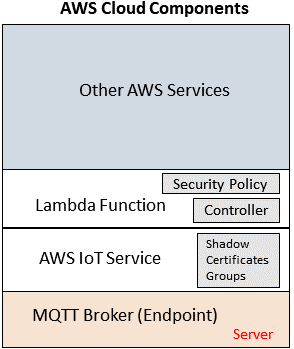} \label{fig-arch1}}
  \caption{AWS Cloud Assisted Prototype Architecture}
  \label{sys-arch}
\end{figure*}

\cvac~is an attribute-based access control model which satisfies fine-grained authorization needs of dynamic, location oriented and time sensitive services and applications in cloud assisted smart cars ecosystem. The model supports personalized privacy controls by utilizing individual user policies and attributes, along with dynamic groups assignment. Our model assumes that the information and attributes shared by source and object entities are trusted, for instance, location coordinates sent by a car are correct, and uses this shared information to make access and notification decisions. How to ensure that the information is from a trusted source or is correct is out of the scope of this work.

\section{\cvac~Model Enforcement in AWS} \label{use-case}

In this section, we present a proof of concept demonstration of \cvac~model by enforcing a use case of smart cars using AWS IoT service \cite{iot}. The implementation will demonstrate how dynamic groups assignment and multi-layer authorization policies required in connected vehicle ecosystem can be realized in AWS. We have used simulations to reflect real connected smart vehicles, however, it does not undermine the plausibility, use and advantage of our proposed model as further elaborated in following discussion. It should be noted that no long term vehicle data including real-time GPS coordinates are collected in central cloud, which mitigates user privacy concerns and encourages wide adoption of the model.

\begin{figure}[t]
\centering
\includegraphics[scale=.50]{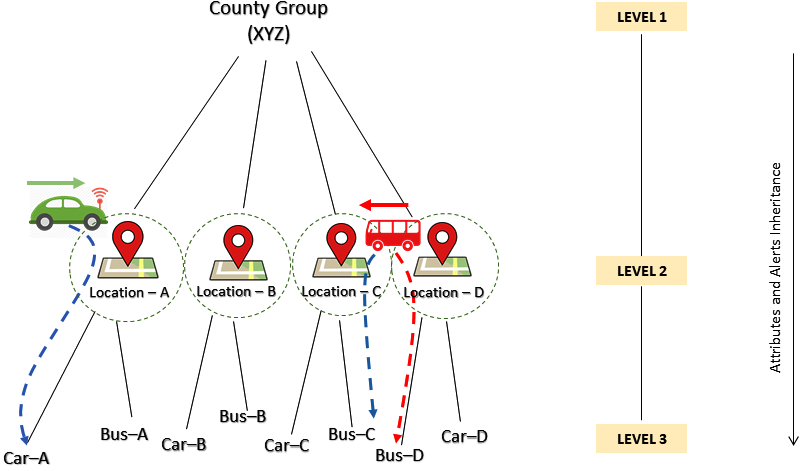}
\caption{Three Level Groups Hierarchy Defined in AWS Implementation}
\label{fig group-hier}
\end{figure}

\subsection{System Architecture}

Figure \ref{sys-arch} shows the overall system architecture along with different components used to implement prototype. Vehicles and infrastructure smart devices are simulated as virtual machines with a MQTT client, sending MQTT payload to the central broker provided by the AWS IoT cloud platform as shown in Figure \ref{fig-arch}. AWS IoT provides a custom endpoint that allows to connect devices to AWS IoT, where each devices have a REST API available at the endpoint. MQTT broker provided by AWS IoT, enables clients (devices) to publish and subscribe to their reserved and secure topic to get and publish messages from other smart entities via the cloud. These reserved topics \cite{reserved-topics} enable device to update, get or delete the state information of the device shadows \cite{shadow-topics} created in the AWS IoT service. Publishing and subscribing to these topics need authorization \cite{bhatt2017access} which ensures that only allowed devices are able to communicate through the cloud. AWS Lambda function \cite{lambda} is an event-driven serverless platform service to run code, and is used to implement, enforce (PEP- Policy Enforcement Point) and decide (PDP - Policy Decision Point) \cite{hu2013guide} the attribute based security polices defined in the system. AWS Lambda function is also used to implement our proposed smart-car controller which helps to assign moving vehicles to dynamic groups and enable attributes and alerts inheritance. Figure \ref{fig-arch1} shows details of AWS cloud components, reflecting where device shadows \cite{shadow}, certificates \cite{certs} and groups \cite{groups} are created in AWS IoT service and MQTT broker acting as a server, providing a client server architecture with IoT devices simulated in the system.

\subsection{Description of Use Cases}
Location based alerts and notifications are important in smart cars applications and motivate our use cases. We will build upon our defined group hierarchy in AWS shown in Figure \ref{fig group-hier}. Our implementation will enforce access controls and notification relevance in following use cases:

\noindent
\textbf{Deer Threat Notification} - Smart infrastructure in the city can sense the surrounding environment and notify group(s) regarding the change. In this use case, a motion sensor senses deers in the area and changes Deer\_Threat attribute of location group to ON which in-turn sends alerts to all member vehicles in that location. Similar, implementation can be done in case of accident notification, speed limit warning or location based marketing.

\noindent
\textbf{Car-Pooling} - A traveller needs a ride to Location-A. Using a mobile application, he sends car-pooling requests to vehicles in his vicinity which are heading to the destination location asked by the traveller. The request is received by AWS cloud, which computes location and appropriate groups based on the coordinates of the requester, to publish notifications to nearby cars. All the members of group Car-A, B, C or D can get the request, but some cars may not want to be part of car-pooling, or do not want some requestors to join them because of ratings. User policies will be also checked before a driver is notified of likely car-pool customer.

\begin{figure}[t]
\centering
\includegraphics[scale=.45]{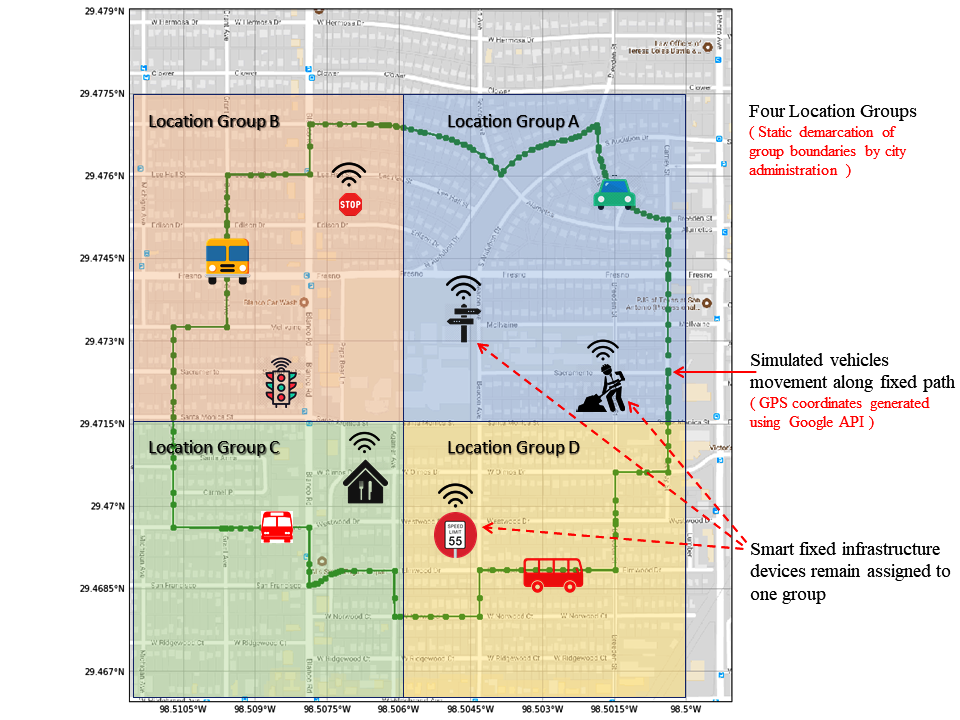}
\caption{Simulated Vehicles Fixed Path and Static Location Groups Demarcation}
\label{fig gps}
\end{figure}

\subsection{Prototype Implementation}
 AWS implementation of our model in these use-cases involves two phases: administrative phase and operational phase. Administrative part involves creation of groups hierarchy, dynamic assignment of moving vehicles to different location and sub-groups, attributes inheritance from parent to child groups and to group members, and attributes modification of entities. Operational part covers how groups are used to scope down the number of vehicles who receive messages or notifications from different sources. Both these phases involve multi-layer access control polices. We created an ABAC policy decision (PDP) and enforcement point (PEP) \cite{hu2013guide}, and implemented our external policy evaluation engine which is hooked with AWS to enable attribute-based authorization.

\begin{figure}[t]
\centering
\captionsetup{justification=centering}
\includegraphics[width=1.0\textwidth]{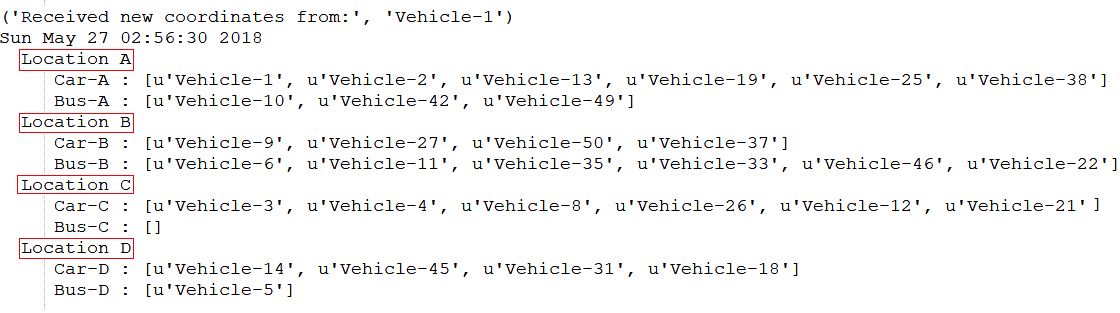}
\caption{Snapshot of Table Showing Dynamic Groups and Associated Connected Vehicles at One Point of Time \\ (Table keeps on updating as vehicles move)}
\label{fig dg}
\end{figure}

\begin{figure}[t]
\centering
\includegraphics[scale=.65]{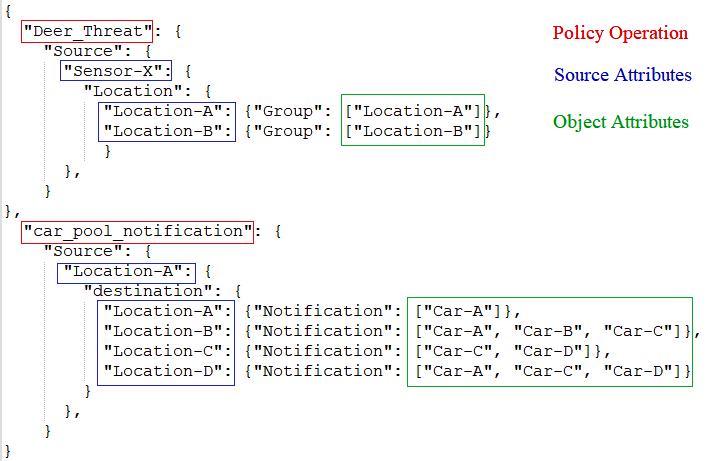}
\caption{Snippet of Attribute Based Policies Implemented in AWS}
\label{fig-policy}
\end{figure}

\noindent
\sloppy
\textbf{Administrative Phase:} We defined a group hierarchy in AWS as shown in Figure \ref{fig group-hier}. In this three level hierarchy, County-XYZ is divided into four disjoint Location-A, B, C and D groups, with each having Car and Bus subgroups for vehicle type car or bus. We created 50 vehicles and simulated their movement using a python script which publishes MQTT message to shadows of these vehicles with current GPS coordinates (generated using Google API \cite{api}) iterated over green dots shown in Figure \ref{fig gps}. The area was demarcated into four locations and a moving vehicle belongs to a subgroup in one of these groups whereas fixed sensor devices remained assigned to one group only. Assuming current location of Vehicle-1 as Location-D, and it publishes MQTT message with payload:
 \begin{itemize}[leftmargin=*]
  \item[] \texttt{\{"state": \{"reported": \{"Latitude": "29.4769353","Longitude":"-98.5018237"\}\}\}}
\end{itemize}
to AWS topic: \texttt{\$aws/things/Vehicle-1/shadow/update}, its new location changes to Location-A and since we defined the vehicle type as car, it is assigned to Car-A group under Location-A as shown by snapshot in Figure \ref{fig dg}. This table keeps on updating as the vehicles move from one location to another, based on their current GPS coordinates sent to the central cloud periodically along with other relevant attributes. Such change ensure that the alerts and notifications received by these vehicles are relevant to their current location. Both attributes, vehicle type and current coordinates of vehicle, are used to dynamically assign groups, which is important in moving smart vehicles. These functionalities are implemented as a stand alone service (can be enforced as a Lambda service \cite{lambda} function) using Boto \cite{boto} which is the AWS SDK for Python. Further, in case of deer threat notification use-case, we simulated a location-sensor which senses deers in the area and updates the attribute `Deer\_Threat' of location group to `ON' or `OFF', which is then notified to all members of location and its subgroups. An attribute-based policy is defined to control which sensors are allowed to change the `Deer\_Threat' attribute of location groups. Figure \ref{fig-policy} shows the snippet of policies implemented in our prototype. The JSON format policy file defines a set of policies for two operations: one for Deer\_Threat and another for car\_pool\_notification, as marked by red box. The blue box specifies the attributes of source, also known as initiator of operation request, whereas the green box specifies the attributes of target object to which the action is requested.
As shown in Figure \ref{fig-policy}, our defined policy for Deer\_Threat operation checks that a motion sensor with name = `Sensor-X' and currently member of group Location-A can update the value of attribute Deer\_Threat for location group Location-A only, and if sensor is relocated to Location-B it can update same attribute for Location-B group only. This policy ensures that the sensor must be in that location group for which it is updating Deer\_Threat attribute, which is needed security requirement as we don't want adversaries to remotely change attributes and trigger unwanted alerts for vehicles.


 A moving vehicle updates its coordinates to AWS shadow service, which along with attributes of vehicles and location groups determines if the vehicle can be member of the group using our external enforcement service.
If authorization policy allows vehicle to be a member of group, the vehicle and group is notified and vehicle inherits all attributes of its newly assigned group. Similarly, if attribute `Deer\_Threat' of group is allowed (by authorization policy) to be changed by the location sensor, the new values are propagated to all its members. We implemented attribute inheritance from parent to child groups through our service using \texttt{update\_thing\_group} and \texttt{update\_thing} methods. In our use-case attributes inheritance exist from Location-A to all both subgroups Car-A and Bus-A, and to vehicles in Car-A and Bus-A. Therefore, when attribute `Deer\_Threat' is set to ON in group Location-A, its new attributes using Boto \texttt{describe\_thing\_group} command are:
\begin{itemize}[leftmargin=*]
  \item[] \texttt{\{`Center-Latitude': `29.4745', `Center-Longitude': `-98.503', }
  \item[] \texttt{ \;\;\;\;\;\;\;\;\;\;\;\;\;\;\;\;\;\;\;\;\;\;\;\;\;\;\;\;\;\;\;\;\;\;\;\;\;\;\;\;\;\;\;\;\;\;\;\;\;\;\;\;\;\;\;`Deer\_Threat': `ON'\}}
\end{itemize}
This inherits the attributes to Car-A child group whose effective attributes will now be:
\begin{itemize}[leftmargin=*]
  \item[] \texttt{\{`Center-Latitude': `29.4745', `Center-Longitude': `-98.503', }
  \item[] \texttt{\;\;\;\;\;\;\;\;\;\;\;\;\;\;\;\;\;\;\;\;\;\;\;\;\;\;\;\;\;\;\;\;\;\;\;\;\;\;\;\;\;\;\;\;\;\;\;\;\;\;\;\;\;\;\;\;\;`Deer\_Threat': `ON', `Location': `A'\}}
\end{itemize}
As shown in Figure \ref{fig dg}, both Vehicle-1 and Vehicle-2 are members of Car-A sub-group, therefore, the effective attributes of Vehicle-2 are:
\begin{itemize}[leftmargin=*]
  \item[] \texttt{\{`Center-Latitude': `29.4745', `Center-Longitude': `-98.503', }
  \item[] \texttt{\;\;\;\;\;\;\;\;\;\;\;\;\;\;\;\;\;\;\;\;\;\;\;\;\;\;\;\;\;\;\;\;\;\;\;\;\;\;\;\;\;\;\;\;\;\;\;\;\;\;\;\;\;\;\;\;\;`Deer\_Threat': `ON', `Location': `A',}
  \item[] \texttt{\;\;\;\;\;\;\;\;\;\;\;\;\;\;\;\;\;\;\;\;\;\;\;`Type': `Car', `VIN': `9246572903752', `thingName': `Vehicle-2'\}}
\end{itemize}
where \texttt{`Center-Latitude',`Center-Longitude',`Deer\_Threat'} and\texttt{`Location'} are inherited attributes from the member group  Car-A and \texttt{`Type', `VIN'} and \texttt{`thingName'} are Vehicle-2 direct assigned attributes. Similar attributes inheritance is witnessed for Vehicle-1 and other vehicles.

\begin{figure*}[t]
\centering
\includegraphics[scale = .48]{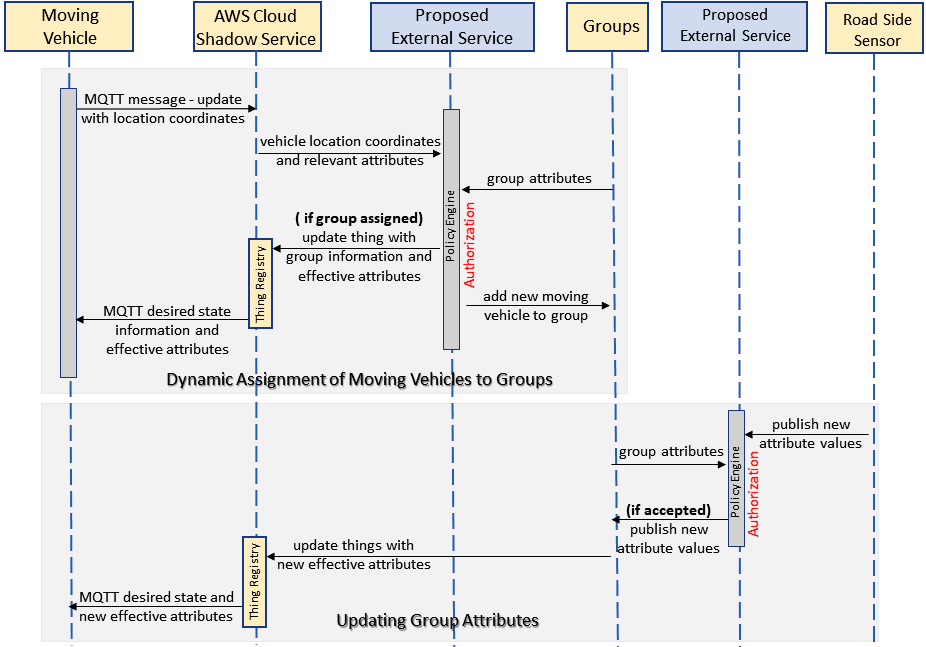}
\caption{Sequence Diagram for Dynamic Groups and Attributes Assignment in AWS}
\label{fig-seq1}
\end{figure*}

 The complete sequence of events performed in AWS along with our stand-alone service for the administrative phase is shown in Figure \ref{fig-seq1}. A moving vehicle sends MQTT message with location coordinates to its reserved topic in the AWS cloud shadow service. Our external stand-alone service, checks the location and attributes of vehicle together with group attributes to determine if the vehicle is within the range of group and can be dynamically assigned. If the vehicle becomes member of the group, its virtual shadow inherits attributes of its member group which are then updated for the real physical vehicle using thing registry. Once a vehicle becomes member of group, any change in the attributes of associated group results in attribute update for the member vehicles. As shown in the lower part of Figure \ref{fig-seq1}, a road side sensor publishing a new value for the attribute, the security policy implemented in the cloud will be checked by the proposed service. If the policy allows the requested change, the values are updated for the members via thing registry and shadow service. This explains the sequence of steps for administrative phase of our prototype.

\noindent
\sloppy
\textbf{Operational Phase:}
In this phase, attribute-based policies are used to restrict service and notification activities which may require single or multi-level policies along with user preferences. In car-pooling use case, we defined policies to restrict notifications to only a subset of relevant vehicles in specific locations. We simulated requestor in AWS needing car-pool. It has attribute `destination' with value in Location-A, B, C or D. Requestor sends current and destination location as MQTT message to AWS topic \texttt{\$aws/things/Requestor/shadow/update}
which based on these attributes determine subgroups to which service requests is sent.
 \begin{itemize}[leftmargin=*]
  \item[] \texttt{\{"state": \{"reported": \{"policy": "car\_pool\_notification",}
  \item[] \texttt{ \;\;\;\;\;\;\;\;\;\;\;\;\;\;\;\;\;\;\;\;\;\;\;\;\;\;\;\;\;\;\;\;\;\;\;\;\;\;\;\; "source": "Location-A",\;\;\;\;\;\;}
  \item[] \texttt{\;\;\;\;\;\;\;\;\;\;\;\;\;\;\;\;\;\;\;\;\;\;\;\;\;\;\;\;\;\;\;\;\;\;\;\;\;\;\;\;\;\; "destination": "Location-B"\}\}\}}
\end{itemize}
The policy for \texttt{car\_pool\_notification} operation (shown in Figure \ref{fig-policy}) suggests that if current location of source requestor is `Location-A' and destination location is somewhere in `Location-A' then all members of sub-group Car-A should be notified. Similarly, if the destination attribute is Location-B, then all members of Car-A, Car-B and Car-C needs notification. In our use-case, all members of these sub-groups are notified. The policy restricts the number of vehicles which will be requested as compared to all vehicles getting irrelevant notification (as they are far from the requestor or are not vehicle type car) and illustrates the importance of location-centric smart car ecosystem. Similarly, location-based marketing can be restricted and policies can be defined to control such notifications.

\begin{figure*}[t]
\centering
\includegraphics[scale = .48]{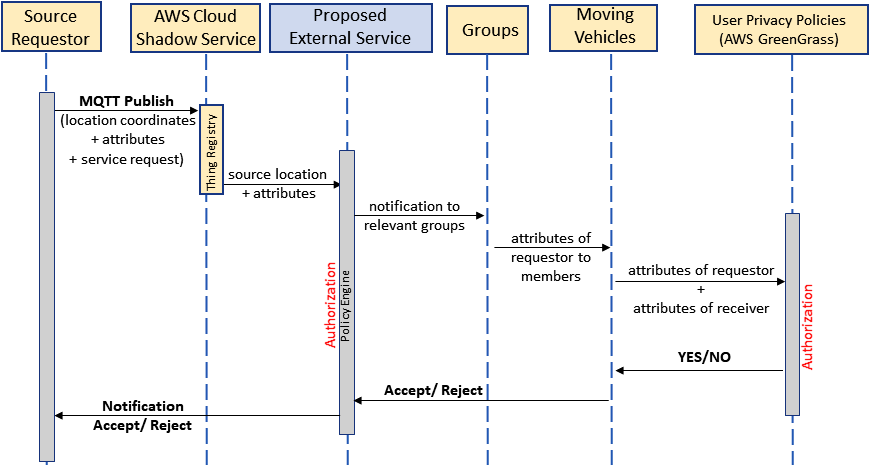}
\caption{Sequence Diagram for Attributes Based Authorization in AWS}
\label{fig seq2}
\end{figure*}

User privacy policies take into effect once the subset of vehicles is calculated. These policies encapsulate user preferences, for instance, in car pooling a particular driver is not going to the destination requested by the requestor in his request or a driver do not want restaurant advertisements, therefore such notifications will not be displayed on his car dashboard. These local policies are implemented using AWS Greengrass \cite{grass} which allows to run local lambda functions on the device (in our case a connected vehicle) to enable edge computing facility, an important requirement in real-time smart car applications and enforce privacy policies. Once accepted by drivers, a SNS (AWS Simple Notification Service) \cite{sns} message can be triggered for requestor from accepting vehicles along with name and vehicle number. The sequence of events for car-pooling activity and multi-layer authorization policies together with user personal preferences is shown in Figure \ref{fig seq2}. As can be seen, the source sends MQTT message with location, attributes and type of service request to the shadow service in the cloud. Once the proposed service (encapsulating the policy decision and enforcement engine) checks the request against the policies defined and enforce decision, notifications are sent to relevant groups and then to member vehicles. Second layer of policy enforcement is done at the individual vehicles. These user privacy policies can be implemented using AWS Greengrass \cite{grass}, where the final authorization check is done. If the user preference policy rejects such notifications, central cloud service is notified and subsequently the request from the source user has been denied.



Our proposed external service to implement ABAC policy decision and evaluation helps achieve fine grained authorization needed in smart cars ecosystem. The implementation also demonstrates dynamic groups assignment based on mobile vehicle GPS coordinates and attributes along with groups based attributes inheritance which offer administrative benefits in enforcing an ABAC model. In this entire implementation, no persistent data from moving vehicles is collected or stored by the central authority hosted cloud which reaffirms its privacy preserving benefits. Note that the use-cases discussed to enforce \cvac~are not real-time and can bear some latency due to the use of cloud infrastructure. Although our \cvac~enforcement in AWS reflects its use for cloud based applications, we believe similar model can also be implemented in edge (or fog) systems as well to cater more real-time use-cases.


\subsection{Performance Evaluation Metrics}

We evaluated the performance of our proposed \cvac~model in AWS and discuss metrics to reflect the impact of our stand-alone external service to have security enhanced smart-car ecosystem. To simulate the environment, we have simulated 50 moving vehicles and used our smart-car controller to randomly spread them across four location sub-groups pre-defined in the system as shown in Figure \ref{fig gps}. In our evaluation, we provide two types of metrics for both the use-cases, the first metric elaborates the policy enforcer execution time for the security policies defined in Figure \ref{fig-policy}, and the second metric provides the comparison of when no policies were used in the system against our implemented ABAC policies. Table \ref{tab-perform} describes our external policy engine execution time for deer-threat and car-pool use-cases. This time (in milliseconds) primarily shows how long it takes to evaluate the implemented policies after action requests for different operations are received by the cloud implemented policy engine. The table aggregates the policy evaluation time against number of action requests, for example, the total time it takes to evaluate the policy for 10 random car-pool requests is 0.0922 ms. Clearly, the engine is very efficient and has minimal impact when used in cloud assisted smart-cars system.

\begin{table}[]
\caption{Attributes Based Policy Enforcement Time (in Milliseconds)}
\renewcommand{\arraystretch}{1.3}
\label{tab-perform}
\begin{tabular}{|c|c|l|c|l|}
\hline
\cellcolor[HTML]{FFFFC7}                                                                                       & \multicolumn{4}{c|}{\cellcolor[HTML]{FFFFC7}Policy Enforcer Execution Time}        \\ \cline{2-5}
\multirow{-2}{*}{\cellcolor[HTML]{FFFFC7}\begin{tabular}[c]{@{}c@{}}Number of Action \\ Requests\end{tabular}} & \multicolumn{2}{c|}{\textbf{Deer-Threat}} & \multicolumn{2}{c|}{\textbf{Car-Pool}} \\ \hline
10                                                                                                             & \multicolumn{2}{c|}{0.0813}               & \multicolumn{2}{c|}{0.0922}            \\ \hline
20                                                                                                             & \multicolumn{2}{c|}{0.1551}               & \multicolumn{2}{c|}{0.2003}            \\ \hline
30                                                                                                             & \multicolumn{2}{c|}{0.2369}               & \multicolumn{2}{c|}{0.2872}            \\ \hline
40                                                                                                             & \multicolumn{2}{c|}{0.3150}               & \multicolumn{2}{c|}{0.3953}            \\ \hline
50                                                                                                             & \multicolumn{2}{c|}{0.3903}               & \multicolumn{2}{c|}{0.5196}            \\ \hline
\end{tabular}
\end{table}

\begin{figure*}[t]
\centering
\includegraphics[scale = .58]{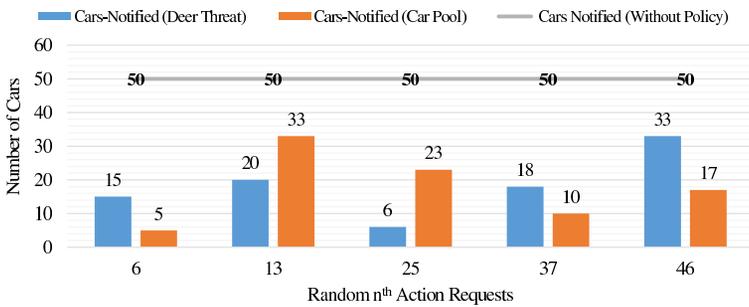}
\caption{Comparing the Scoping and Relevance of Alerts with and without Policy}
\label{fig-notify}
\end{figure*}

Next, we show the impact of how enforcing policies in the system ensures the relevance and scope of alerts received by smart cars. One of the major push and advantage for smart and connected vehicles is to have on-board advertisements and alerts to offer convenience and safety to the drivers. But at the same time, drivers also do not want to be bothered by notifications which are completely irrelevant and extraneous like receiving a coupon from a restaurant which is 50 miles away. Therefore, to ensure such incidents do not pester and divert drivers attention, our proposed ABAC policies can be helpful. Figure \ref{fig-notify} shows the number of vehicles notified for deer threat and car-pool notifications with and without the policy. Without the policy, irrespective of the vehicle location or the driver personal preferences all the vehicles in the system are notified (which in our case is 50) when a random request is generated. However, when the cloud based policies are enforced, it ensures that only vehicles to which the notifications are relevant are alerted. For example, in Figure \ref{fig-notify}, on 25\textsuperscript{th} car-pool request only 23 vehicles were notified for the request as compared to all vehicles even when one would have been 20 miles away from the requestor. These subset of vehicles is calculated based on the number of cars which are in the location groups which are near to the originating source, or the manner in which attribute based polices are defined by the administrator. Similar results can be shown for deer-threats alerts, where only cars in the proximity of sensed deers are notified. It must be noted that, in Figure \ref{fig-notify} we have clubbed together the notified cars for both the use-cases, however, the n\textsuperscript{th} request for car-pool is completely separate from the n\textsuperscript{th} request for deer-threat case. The prime motive of this metric is to reflect how the policies enable notification relevance and scoping of target vehicles.

\begin{figure*}[t]
  \centering
  \subfigure[Deer Threat Use Case]{\includegraphics[scale=0.36]{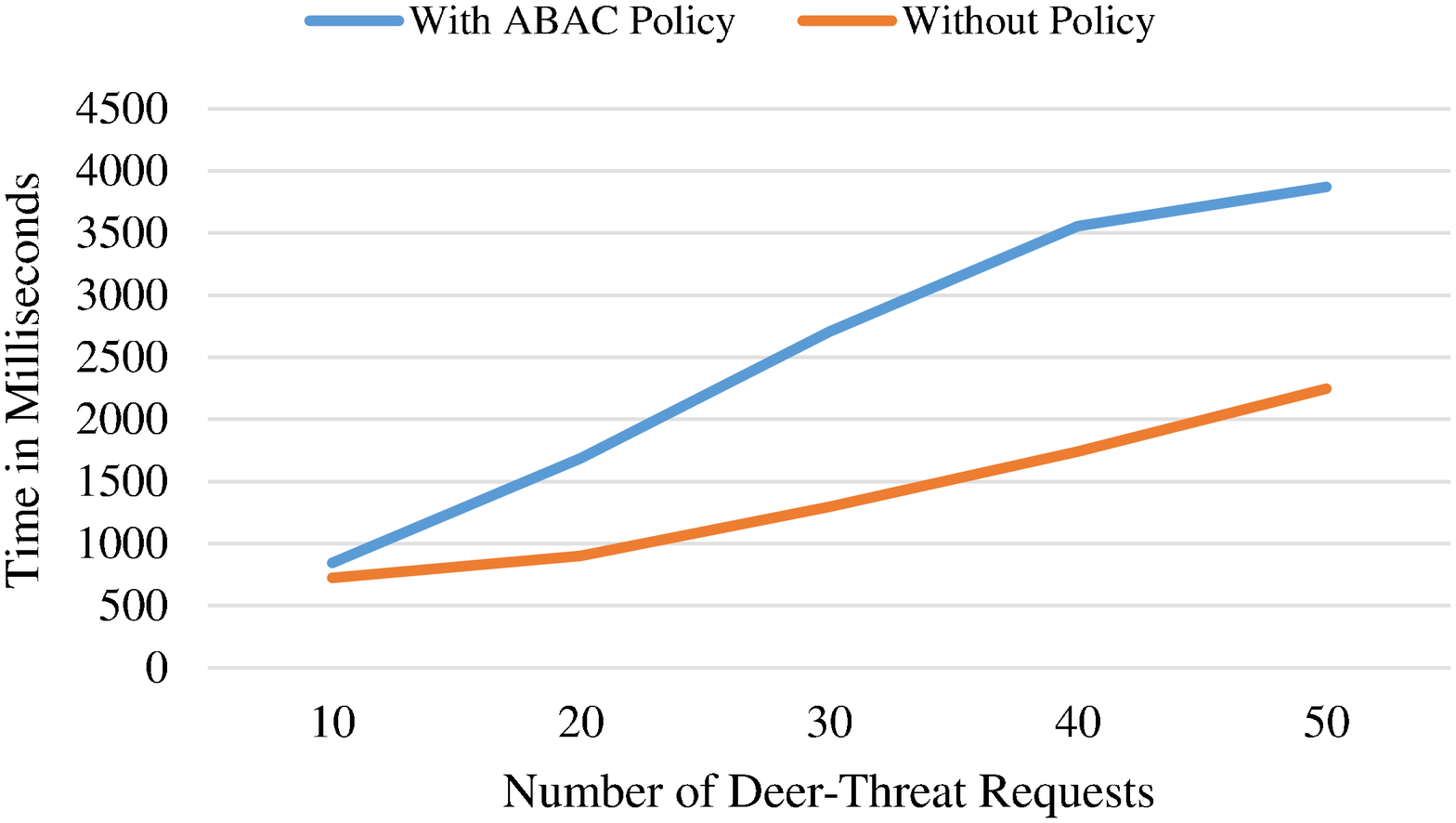}\label{fig-deer}}\quad
  \subfigure[Car Pool Use Case]{\includegraphics[scale=0.36]{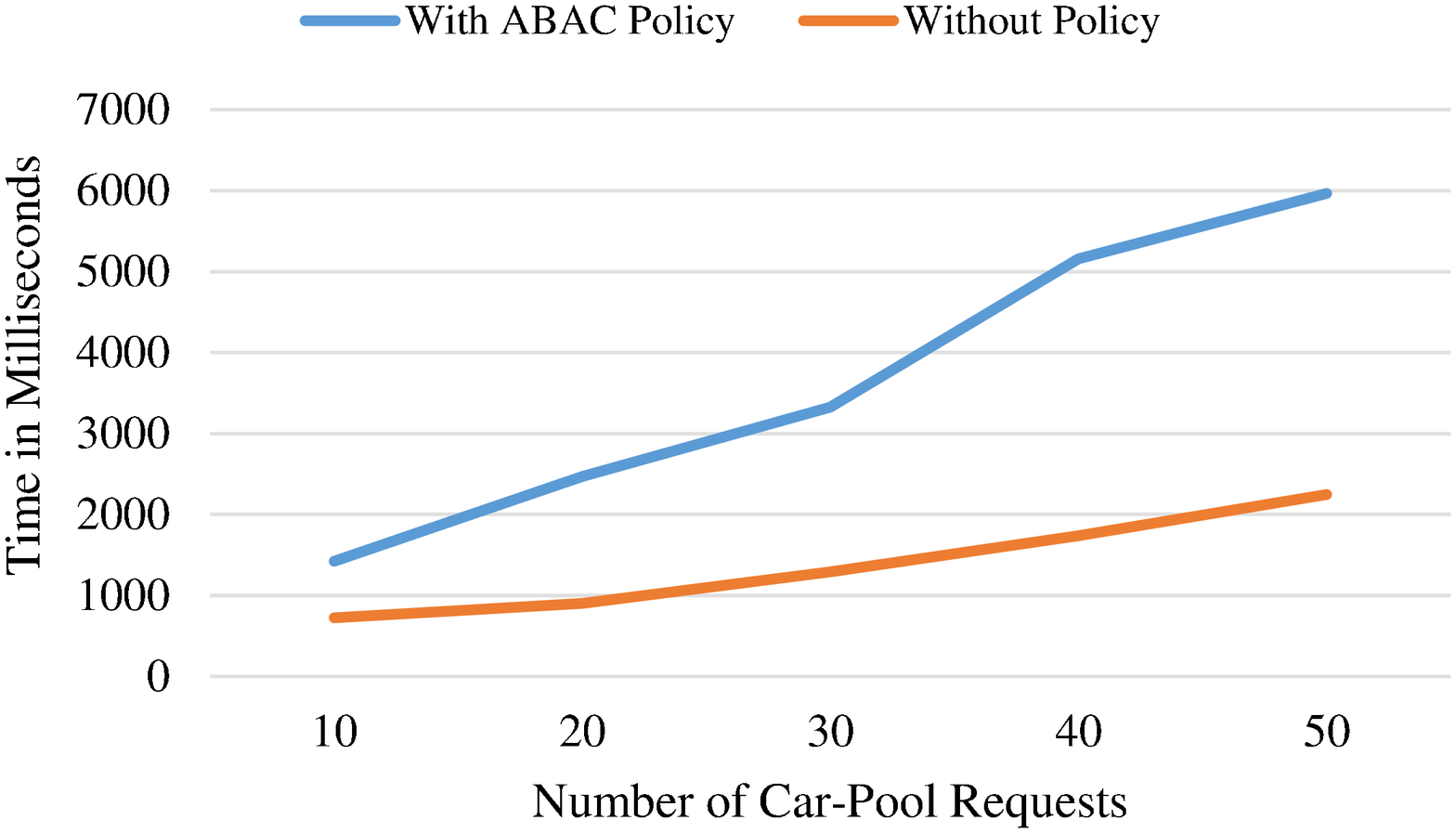}\label{fig-cp}}
  \caption{Performance Comparison with and without ABAC policy}
  \label{fig-graph}
\end{figure*}

The performance graph shown in Figure \ref{fig-graph} compares the execution time when no policy is executed (orange line) against implemented ABAC policy (blue line) for deer-threat and car pool use-cases. Since the main focus of this experiment is to measure the impact of the proposed ABAC policy based security solution, this metric evaluates the time it takes to calculate the list of vehicles to be notified with and without the policy. The X axis is each graph shows the total number of execution requests i.e. the number of times deer-threat (Figure \ref{fig-deer}) or car-pool (Figure \ref{fig-cp}) notifications are generated and Y axis denote the total time (in milliseconds) from the moment the access or notification request is received by the Lambda function in cloud till the number of vehicles to be notified is calculated. Since, in our experiments the policy (shown in Figure \ref{fig-policy}) definition for each access requests in both deer-threat and car-pool are near identical, we observe that the number of access requests increase the number of times the policy is evaluated and so its total evaluation time also increases. Minor variations in the orange and blue lines (in graphs Figure \ref{fig-graph}) are because of AWS API endpoint calls being made from the Lambda function to calculate the number of vehicles notified in both the cases, which can change based on the internal communication technologies used by the cloud services. Our proposed external policy engine does have some impact on the performance (as shown with blue line) as opposed to no policy when used with number of vehicles. However, we believe when used in city wide scenario this time will be overshadowed by cloud assisted notification time to all vehicles against a subset of vehicles provided by the policy evaluation engine. Our model and the prototype implementation of use-cases are focused to ensure service relevance to moving vehicles on road which is well achieved even with a little tradeoff.

The prototype implementation and performance metrics in Section \ref{use-case}, provide a comprehensive understanding of how the capabilities of cloud are used to provide a secure and privacy aware smart car environment. In this implementation, we illustrated how attributes based polices can be implemented in the system, and their application to ensure fine grained access control and activity centric authorization in smart cars. The most important advantage of using cloud based security service is the `infinite' capabilities and auto-scaling provided by the cloud which can help to cater a large number of smart vehicles in city wide geography. As mentioned earlier, their are associated administrative operations in this proposed solution like demarcating location group boundaries, setting group hierarchy, administering and modifying attributes of groups, defining attributes based security policies etc. which must be done in a diligent manner by the city administration to fully realize the potential of proposed secure cloud assisted smart cars ecosystem.

It is considered that a practical smart city transportation scenario will have hundreds and thousands of moving cars (and other connected entities) associated to cloud (or fog infrastructures) interacting and initiating alerts for entities. Although a detailed performance evaluation is very desirable by having large sets of real moving vehicles, we believe that our proof of concept in AWS is to showcase the practical viability, application and use of fine grained attribute based security policies in context of smart cars ecosystem, without the need to capture large set of data points from real world traffic scenarios spread across wide geographic area and sizable on-road moving vehicles. Such scaled setting will only stress the entire system without reflecting any change in security policy evaluation. It should be noted that Amazon Web Service (AWS) is just one of the cloud based platforms to realize the proposed model and similar prototype can be implemented in other cloud computing services including Microsoft Azure \cite{azure}, Google Cloud \cite{google-iot} or Openstack \cite{openstack}. The main objective of this paper is to propose the specification and introduction of ABAC policies in a cloud assisted smart cars environment without focusing on any one cloud platform.

\section{Summary} \label{summary}
This research work presents a fine-grained attribute-based access control model for time-sensitive and location-centric smart cars ecosystem. Our model introduces the novel notion of dynamic groups in relation to connected vehicles and emphasizes its relevance in this context. Besides considering system wide authorization policies, this model also supports personal preference policies for different users, which is required in today's privacy conscious world. Several real world use-cases are discussed and a proof of concept implementation of our \cvac~model is shown in Amazon Web Services (AWS) cloud platform. We created a smart car controller to demonstrate how moving vehicles can be dynamically assigned to location and sub-groups defined in the system based on the current GPS coordinates, vehicle-type and other attributes, besides the use of attribute based security policies in distributed and mobile connected cars ecosystem. Detailed performance metrics have been evaluated for deer-threat and car-pool use-cases to determine activity access control decision when groups and ABAC policies are used against when no security policies are available. We plan to extend this model to introduce in-vehicle security and built risk aware trust-based models for smart vehicles environment. Further, it is important to introduce location privacy preserving approaches such as homomorphic encryption and other anonymity techniques to complement and extend our model which can mitigate location sharing concerns without effecting its advantages and application. Similar approach for V2X trusted DSRC communication and privacy concerns also need further investigation, which we are currently working in a direction to propose a secure intelligent transportation system.

\bibliographystyle{ACM-Reference-Format}
\bibliography{acmart}
\end{document}